\documentclass[aps,twocolumn,prc,superscriptaddress,showpacs,nofootinbib,floatfix,amssymb,amsfonts,amsmath]{revtex4-1}
\usepackage{graphicx}
\usepackage{dcolumn}
\usepackage{bm}

\usepackage{amsmath}    
\usepackage{amsfonts}   
\usepackage{amssymb}
\usepackage{graphicx}   
\begin{document}

\title{The impact of collective vibrations on quasiparticle states 
of open-shell odd-mass nuclei and possible interference with 
tensor force}

\author{A.\ V.\ Afanasjev}
\affiliation{Department of Physics and Astronomy, Mississippi
State University, MS 39762}

\author{E.\ Litvinova}
\affiliation {Department of Physics, Western Michigan University,
Kalamazoo, MI 49008-5252, USA} \affiliation{National Superconducting
Cyclotron Laboratory, Michigan State University, East Lansing, MI
48824-1321, USA}

\date{\today}

\begin{abstract}

 The impact of collective vibrations on quasiparticle
states of open-shell odd-mass nuclei and their possible interference
with tensor force is investigated. The inclusion of collective 
vibrations and their coupling to single-quasiparticle motion improves 
the description of experimental spectra in such nuclei. 
We found that 
the energy splittings of the single-quasiparticle states,
which are sensitive at the mean field level to the tensor part of 
the effective nucleon-nucleon interaction, are affected by 
quasiparticle-vibration coupling. Both quasiparticle-vibration 
coupling and effective tensor interaction act in the same direction. 
This suggests that effective tensor interaction has to 
be quenched as compared with previous estimates. 
\end{abstract}


\maketitle

\section{Introduction}

The tensor force is known to be one of the important components of
the bare nucleon-nucleon interaction. However, it is still an open
question
whether or not the effective tensor force used in the density
functional theory (DFT) framework for medium-mass and heavy nuclei
keeps a close resemblance with the original bare tensor force
\cite{SS.14}. Moreover, the question of unambiguous signatures of
the tensor force remains open \cite{SS.14}. For example, the fits of
the energy density functionals (EDF's) to bulk nuclear properties
(masses, radii) usually disfavor effective tensor interaction
\cite{SS.14}; note that such fits are standard in the DFT. On the
other hand, some possible indications of the presence of effective
tensor interaction come from the analysis of the spectra of
predominantly single-particle states in odd-mass nuclei; they are
discussed in detail below. Possible manifestations of effective
tensor force in the DFT framework have been recently reviewed in
Ref.\ \cite{SS.14}. Although there is a general consensus that the
tensor component has to be added to the DFT
\cite{FRS.98,OSFGA.05,LBBDM.07,CSFB.07,ZDSW.08,Zet.09,SS.14}, the
question about its strength is still not fully resolved.

 Following Ref.\ \cite{OSFGA.05}, there were attempts to find a
signature of the effective tensor force in the evolution of the
single-particle states along the isotopic or isotonic chains. For
example, the energy differences between the proton $1g_{7/2}$ and
$1h_{11/2}$ quasiparticle states in the Sb ($Z=51$) isotopes 
and the neutron $1i_{13/2}$ and $1h_{9/2}$ quasiparticle states 
in the $N=83$ isotones were considered in Skyrme DFT calculations 
with the SLy5 functional in Ref.\ \cite{CSFB.07}. It was concluded 
that these energy splittings can be reproduced only when the tensor 
component is added to the functional. A similar investigation has 
been performed with the Gogny functionals but only for $N\geq 64$ 
isotopes in Ref.\ 
\cite{OMA.06} and lead to the same conclusion. The studies of the 
Sb isotopes with $N\geq 66$ in Ref.\ \cite{LKSOR.09} within the 
relativistic Hartree-Fock approach showed that the relevant energy 
splittings are only reproduced when the pion tensor coupling of 
half of its standard value is included. However, extensive 
multiparameter minimizations led to the conclusion that 
the optimal fit to bulk properties of infinite nuclear matter and 
spherical finite nuclei is achieved for the vanishing pion field, 
and, thus, no tensor force.

  It is necessary to recognize that these DFT studies miss
important physics related to the fragmentation of the single-particle
states due to coupling with vibrations which is very strong in 
many nuclei. They assume that above mentioned states are of 
single-particle nature despite the fact that their fragmentation is 
supported by numerous experimental observations 
\cite{Mitch.12,117Sb-spec,136Xe,Kay2008216,CHC.68,SP.08}.  
Thus, mean field studies ignore the realistic structure of the wavefunction 
of the states and, as a consequence, the fit of the parameters of 
effective tensor force to their energies can be misleading.

 The proper way to proceed is to take into account particle-vibration 
coupling (PVC) which affects both the wavefunctions and the energies of 
the states of odd spherical nuclei \cite{Litvinova2006_PRC73-044328,LA.11}.
However, this has never been done in the context of the study of the
effective tensor force because such investigation involves open-shell 
nuclei in which the pairing correlations of the superfluid type have to 
be taken into account.

On the other hand, it has been found that in both non-relativistic and 
relativistic frameworks the PVC improves the description of the energies 
and wave functions of the levels with dominant single-particle content 
in odd mass neighbours of doubly magic nuclei. In the 
relativistic framework this was verified in the systematic studies of 
single-particle spectra of nuclei neighbouring to $^{56}$Ni, $^{132}$Sn 
and $^{208}$Pb in Ref.\ \cite{LA.11}  (see Ref.\ \cite{CCSB.14} and references
therein for non-relativistic results). In addition, Ref.\ \cite{LA.11} 
showed for the first time that PVC has an appreciable impact on the 
energy splittings of the spin-orbit and pseudospin doublets.  For
example, PVC decreases the splitting of the $1f_{7/2}-1f_{5/2}$
spin-orbit doublet of $^{56}$Ni by approximately 1.5 MeV. This
doublet has been used in the definition of the strength of the
tensor interaction in Refs.\ \cite{ZDSW.08,Zet.09} where the effects
of the PVC on its magnitude were neglected.

  The question of whether effective tensor force has to be directly 
included for a proper description of the single-particle spectra 
and other physical observables is very important, especially in the 
framework of the CDFT. The absolute majority of very succesful 
applications of CDFT to different physical phenomena  have been 
performed at the Hartree level \cite{VALR.05,NVR.11}. However, the 
inclusion of effective tensor force requires the transition to the 
Hartree-Fock (HF) level \cite{LSGM.07,LKSOR.09}.  Unfortunately, 
this leads to a drastic increase of required computational power, which 
basically restricts the application of the CDFT at the HF level to the 
spherical nuclei. The only existing demonstration of such an application 
to axially deformed nuclei in Ref.\ \cite{EKAV.11} has used small basis 
and showed a requirement for tremendous computational power. In addition, 
one has to mention that for the absolute majority of physical 
observables the global descriptions at the Hartree and Hartree-Fock 
levels are comparable, which definitely indicates that the 
effects of the Fock term are taken into account at the Hartree level in 
an effective way during the fit of functional to experimental data.

 The main goal of this paper is to understand to which degree beyond 
mean-field effects  (such as the particle-vibration coupling) can 
modify the conclusions about the role of tensor interaction
obtained in the mean field studies.
In order to achieve that
we have chosen the observables which are known to be sensitive 
to effective tensor interaction and used for the tuning of its 
strength. These observables are the energy splittings between the 
proton $1g_{7/2}$ and $1h_{11/2}$ states in the Sb ($Z=51$) isotopes 
and the neutron $1i_{13/2}$ and $1h_{9/2}$ states in the $N=83$ isotones.  
These splittings are studied within the relativistic 
quasiparticle-vibration (RQVC) model developed in Ref.\ \cite{L.12}.

   The paper is organized as follows. Section \ref{calc-det}
describes the details of the calculations. Collective excitations
in even-even nuclei are considered in Sec.\ \ref{coll-exc}. The spectra
of $^{116}$Sn and $^{148}$Dy nuclei obtained in the mean field and 
quasiparticle-vibration coupling calculations are compared with 
experimental data in Sec.\ \ref{spectra}. Sec.\ \ref{splittings} 
presents the discussion of the energy splittings of the dominant 
single-particle states sensitive to tensor interaction. Experimental 
and calculated spectroscopic factors of the states under consideration 
are compared in Sec.\ \ref{s-factors}. Finally, Sec.\ \ref{concl} 
summarizes the results of our work.

\section{The details of calculations}
\label{calc-det}

  The RQVC model \cite{L.12} employed in the present paper
is an extension of the PVC model of Ref.\ 
\cite{Litvinova2006_PRC73-044328} and takes into account 
pairing correlations of the superfluid type in the 
quasiparticle-phonon coupling self-energy. 
Compared to some perturbative approaches to the PVC 
\cite{CSB.10,CCSB.14,TDTC.14}, 
which, in fact, remain on the mean-field level and do not include
pairing correlations, our model implies an exact solution of the Dyson 
equation for the single-quasiparticle propagator with a singular 
frequency-dependent self-energy \cite{Litvinova2006_PRC73-044328,L.12} 
and includes quasiparticle-vibration coupling (QVC) and pairing 
on equal footing.  These features of our approach allow a proper 
description of the nuclei of interest. The RQVC
model has been successfully tested on the single-particle spectra
in proton and neutron subsystems of $^{116,120}$Sn, in which it
substantially improves the accuracy of the description of the spectra
as compared with mean field calculations, and applied to
superheavy nuclei \cite{L.12}.

The calculations have been performed with the NL3*  covariant
energy density functional (CEDF) \cite{NL3*};  it is designed 
at the Hartree level and does not (and not able to) include tensor 
interaction. This functional represents one of very few relativistic 
functionals 
the performance
of which has been tested globally with respect of the ground state
observables \cite{AARR.14} and systematically in the local regions
of nuclear chart with respect of other physical observables
such as the energies of one-quasiparticle deformed states in 
actinides \cite{AS.11} and the moments of inertia of even-even and
odd mass nuclei \cite{AO.13}. In the context of this
paper, it is important that NL3* has been successfully used in the
relativistic PVC studies of the spectra of odd spherical nuclei
adjacent to doubly magic nuclei (such as $^{100,132}$Sn) and the
impact of PVC on physical observables \cite{LA.11}.

\begin{figure}[ht]
\centering
\includegraphics[width=8.0cm]{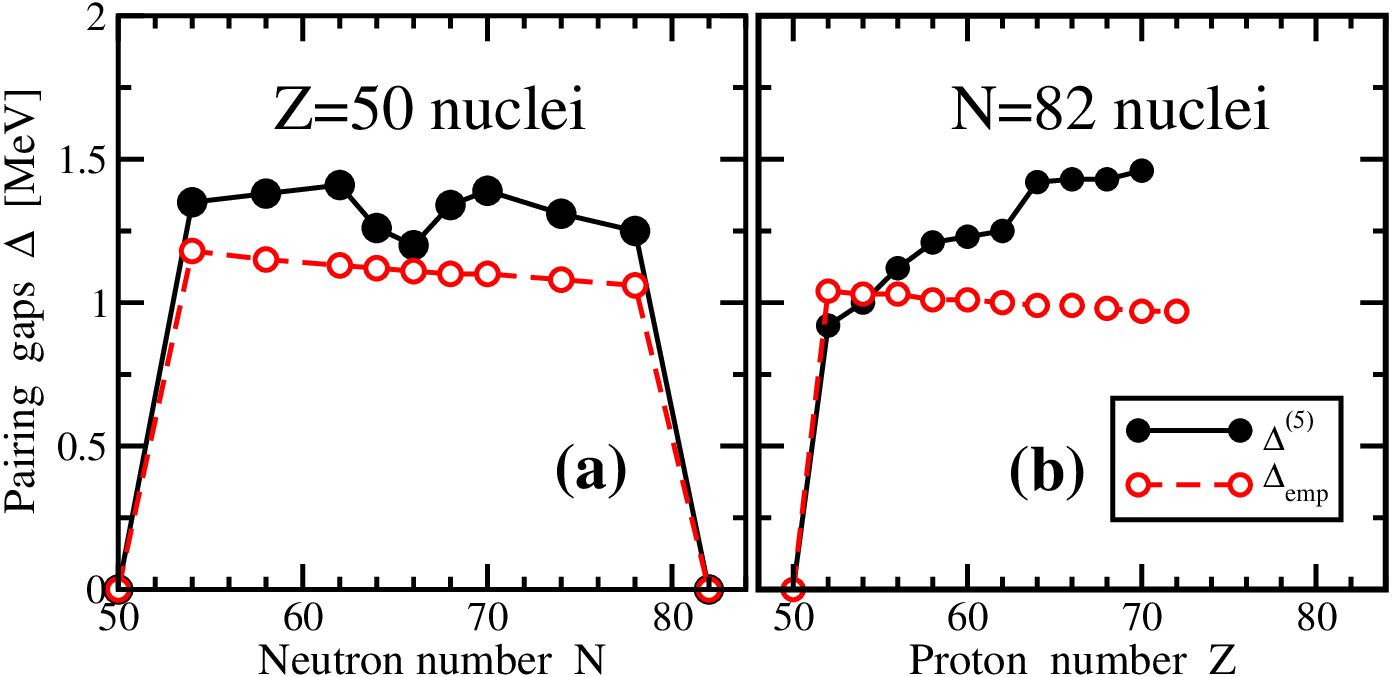}
\caption{ (Color online) Pairing gaps used in the calculations
         as a function of particle number. The $\Delta^{(5)}$ pairing
         gaps are extracted using experimental binding energies from
         Ref.\ \cite{AME2012}.
          }
\label{pairing}
\end{figure}

   For the superfluid pairing correlations, monopole pairing force 
is used in our calculations. In order to study the dependence of the 
results of the calculations on the strength of pairing, two types of 
the parametrizations for the pairing gaps are employed. First, the 
calculations are performed with empirical pairing gap
$\Delta_{emp}=12/\sqrt{A}$ MeV \cite{NilRag-book}, where $A$ stands
for the mass number. Second, the pairing gaps equal to the five-point
indicators $\Delta^{(5)}$ \cite{BRRM.00}
\begin{eqnarray}
\Delta^{(5)}(X) &=& \frac{(-1)^X}{8} [ E(X+2)-4E(X+1) \nonumber \\
                & & + 6E(X) - 4E(X-1) + E(X-2) ]
\end{eqnarray}
defined for each nucleus under study from experimental odd-even mass
staggerings are also used. Here $X$ stands either for proton $Z$ or
neutron $N$ numbers and $E(X)$ is the (negative) binding energy of a
nucleus with $X$ particles. As shown in Ref.\ \cite{BRRM.00}, this
indicator provides the best decoupling from the mean-field effects
in model calculations.
Below, we will refer the corresponding calculations as the 
``$\Delta_{emp}$'' and ``$\Delta^{(5)}$'' ones.  The 
pairing gaps obtained in these calculations are shown in
Fig.\ \ref{pairing}. In the Sn isotopic and $N=82$ isotonic chains,  the $\Delta_{emp}$ 
pairing gaps are close to 1.1 MeV.
The $\Delta^{(5)}$ pairing gaps are larger by roughly 0.2 MeV than the 
$\Delta_{emp}$ ones in the Sn isotopes. In the $N=82$ isotones, the 
difference $\Delta^{(5)}-\Delta_{emp}$ increases from $\sim 0$ MeV at 
$Z\sim 54$ up to $\sim 0.5$ MeV at $Z=70$.  

In the RQVC calculations the phonon space includes normal phonon modes 
of natural parity and it is truncated by the 
angular  momenta of the phonons at $J^\pi=6^+$ and by their frequencies 
at 15 MeV. Further extension of the phonon space does not affect the 
results. Other limitations of the approach are discussed in Ref.\ 
\cite{L.12}.

\section{Collective excitations in even-even nuclei}
\label{coll-exc}

\begin{figure}[ht]
\centering
\includegraphics[width=8.0cm]{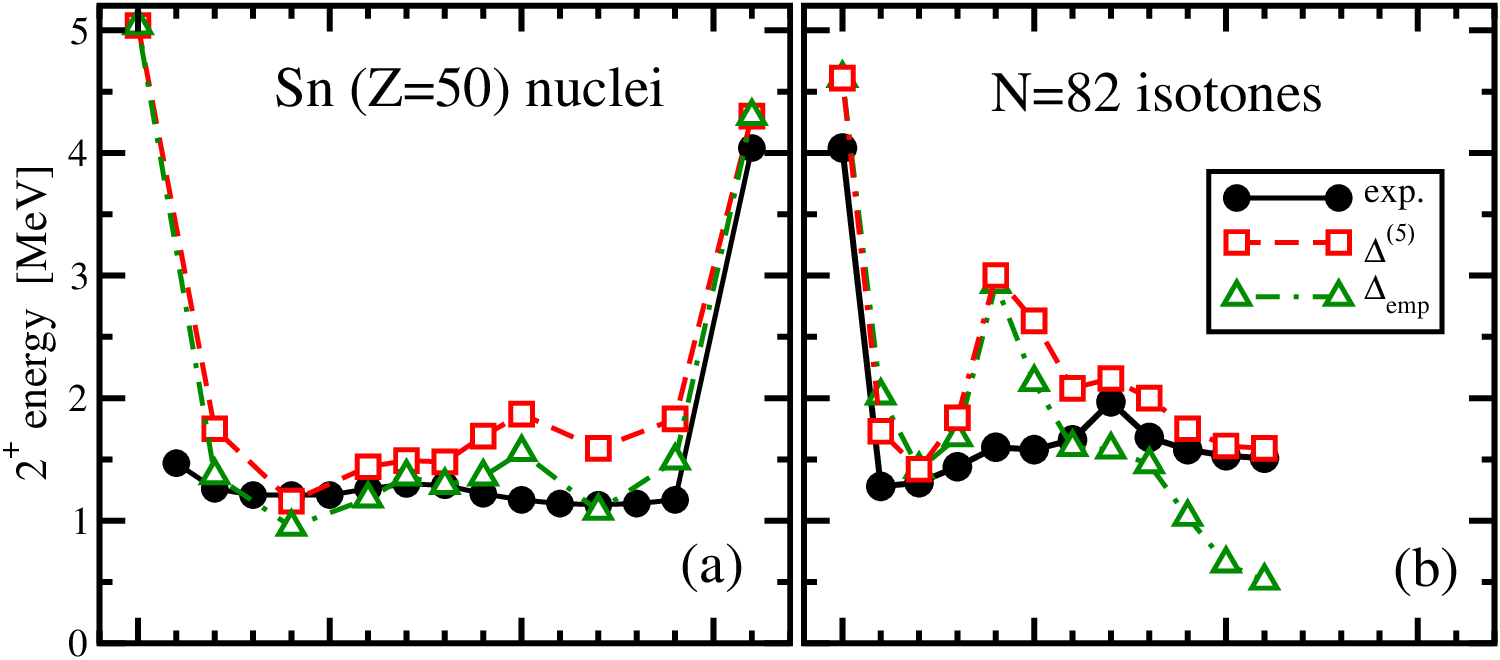}
\includegraphics[width=8.0cm]{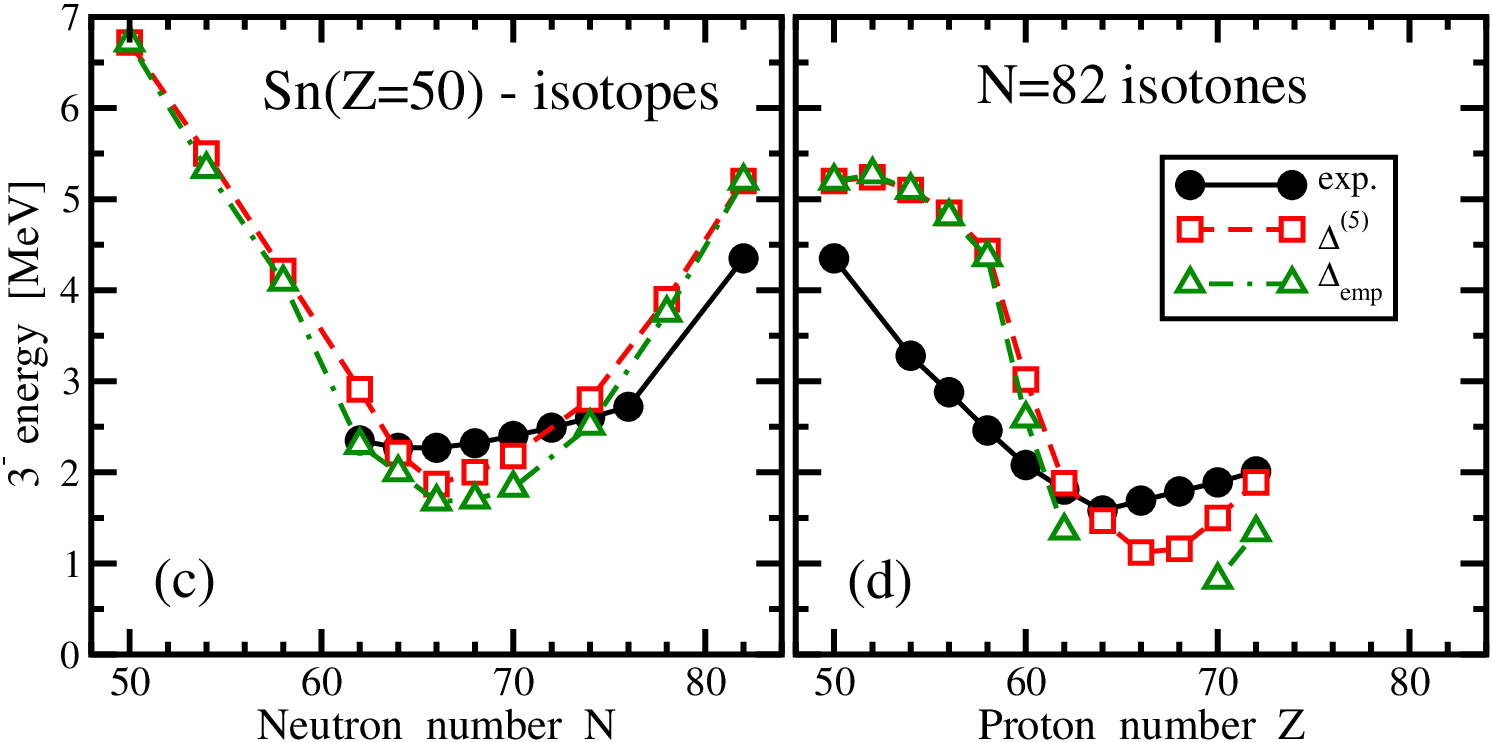}
\caption{(Color online) The excitation energies $E(2^+)$
(top row) and $E(3^-)$ (bottom row) of the lowest $2^+$ and
$3^-$ states in Sn nuclei (left column) and $N=82$ isotones
(right column).  The experimental data are taken from Ref.\
\protect\cite{Eval-data-up}.  Note that for the $N=82$
isotones two-proton drip line is located at $Z=74$ in the
relativistic Hartree-Bogoliubov calculations with the NL3*
CEDF \cite{AARR.14}.}
\label{2+3-energy}
\end{figure}

\begin{figure}[ht]
\centering
\includegraphics[width=8.5cm]{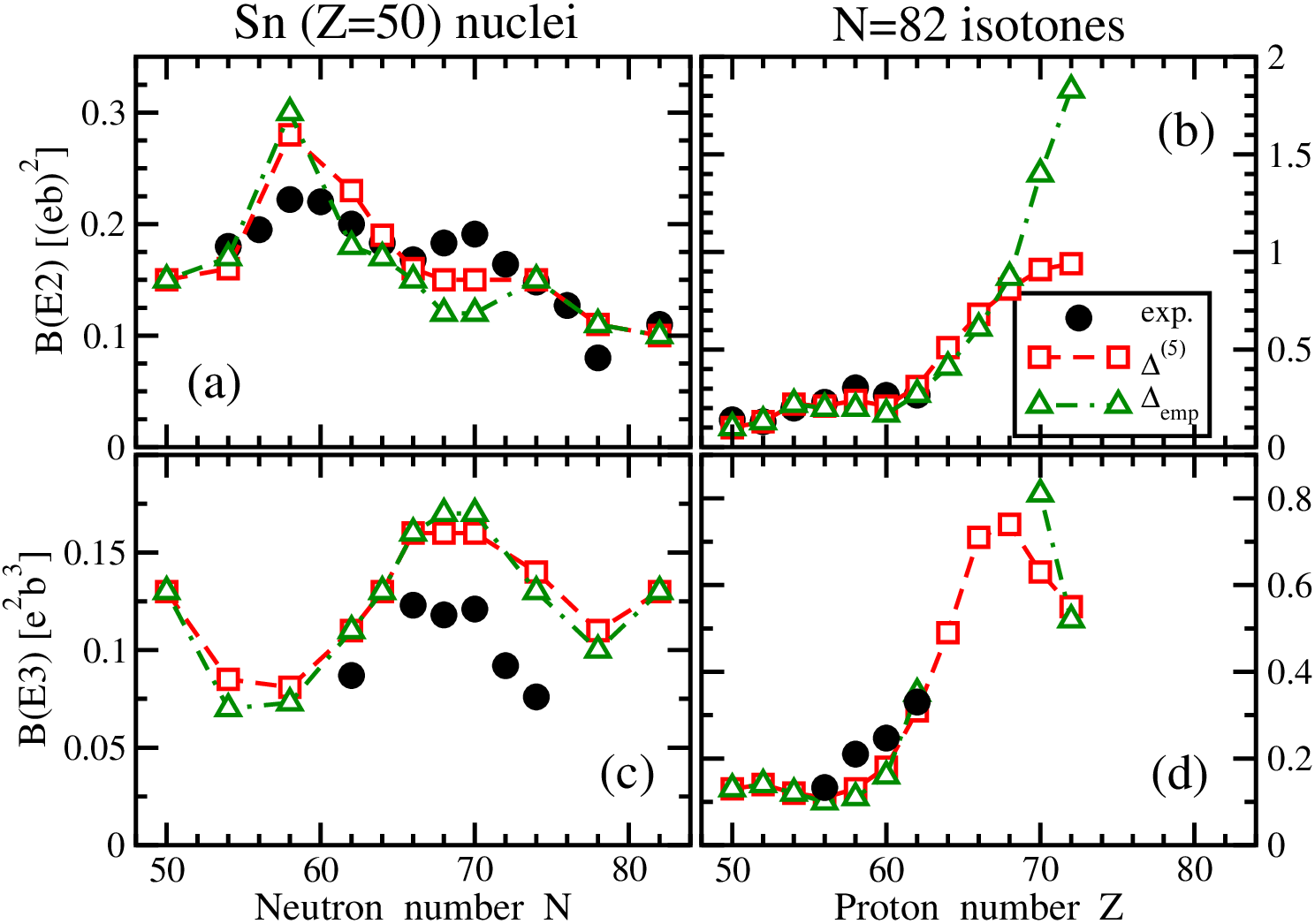}
\caption{(Color online) Reduced transition probabilities
$B(E2; 0^+_{gs} \rightarrow 2^+_1)$ (top row) and
$B(E3; 0^+_{gs} \rightarrow 3^-_1)$ (bottom row)
for Sn nuclei (left column) and $N=82$ isotones (right
column). The experimental data are taken from Refs.\
\protect\cite{Eval-data-up,Bet.14}.}
\label{reducedBE}
\end{figure}

  Before considering the energy splittings of specific
pairs of the states in odd-mass nuclei, it is necessary to
understand how well the excitations energies and the electric
multipole decay rates of the lowest vibrational states in the
Sn and $N=82$ even-even nuclei are reproduced in the Relativistic
Quasiparticle Random Phase Approximation (RQRPA). These
characteristics play a decisive role in the RQVC model: the
nucleonic self-energy is a function of the phonons' frequencies
and quasiparticle-phonon coupling vertices (see, for instance,
Eq.\ (5) of Ref. \cite{L.12}) while the latter are directly related
to the decay rates. Thus, the accuracy of their description defines
the quality of the description of the states with significant
single-particle content in odd-mass nuclei. The excitation energies
of the lowest $2^+$ and $3^-$ states and the corresponding reduced
transition probabilities are shown in Figs.\ \ref{2+3-energy}
and \ref{reducedBE}, respectively. 

The experimental $2^+$ energies of the Sn isotopes are better
reproduced in the ``$\Delta_{emp}$'' calculations, while the $3^-$ ones
are better described in the ``$\Delta^{(5)}$'' calculations (see Figs.\
\ref{2+3-energy}a and c). The experimental reduced $B(E2)$
transition probabilities are slightly better described in the
``$\Delta^{(5)}$'' calculations (Fig.\ \ref{reducedBE}a), while
comparable accuracy of  the description of experimental reduced
$B(E3)$  transition probabilities is achieved in both calculations
(Fig.\ \ref{reducedBE}c). The level of agreement with experiment obtained in the current
calculations is at least comparable with the one obtained in earlier
relativistic \cite{AR.06} and non-relativistic \cite{CTP.12} QRPA
calculations  for the Sn isotopes.

Apart of the $Z=58, 60$ nuclei, the excitations energies of the $2^+$ states
in the $N=82$ isotones are reasonably well reproduced in the model
calculations (Fig.\ \ref{2+3-energy}b).
The results for $Z=58, 60$ are affected by too big shell gap at
$Z=58$, which appear in many CEDF's. It was shown in Ref.\
\cite{LSGM.07} that $\rho$-meson tensor coupling can reduce
the size of this gap.
The results for the reduced $B(E2)$ transition probabilities
are, nevertheless, close to experiment (Fig.\
\ref{reducedBE}b). Experimental excitation energies of the $3^-$
states are overestimated for $Z\leq 58$ (Fig.\ \ref{2+3-energy}d)
because of the same reason as the $2^+$ energies, and only above this
neutron number the results of the calculations come close to
experimental data. It turns out that for the case of
``$\Delta_{emp}$'' pairing the QRPA calculations of the $3^-$
excitations in the $Z=64, 66$ and 68 $N=82$ isotones lead to the
appearance of the Goldstone modes (the stability matrix is
not positively defined). Only for the pairing
strength larger than some critical value the energies and the
B(E3) values of the lowest $3^-$ states have physical magnitudes.
The critical value for the pairing gap is larger than the
``$\Delta_{emp}$'' for the nuclei with $Z=64-68$. Therefore, no
meaningful calculations can be done for them within the RQVC
model.

\section{Spectra of $^{116}$Sn and $^{148}$Dy}
\label{spectra}

\begin{figure}[ht]
\centering
\includegraphics[width=4.2cm]{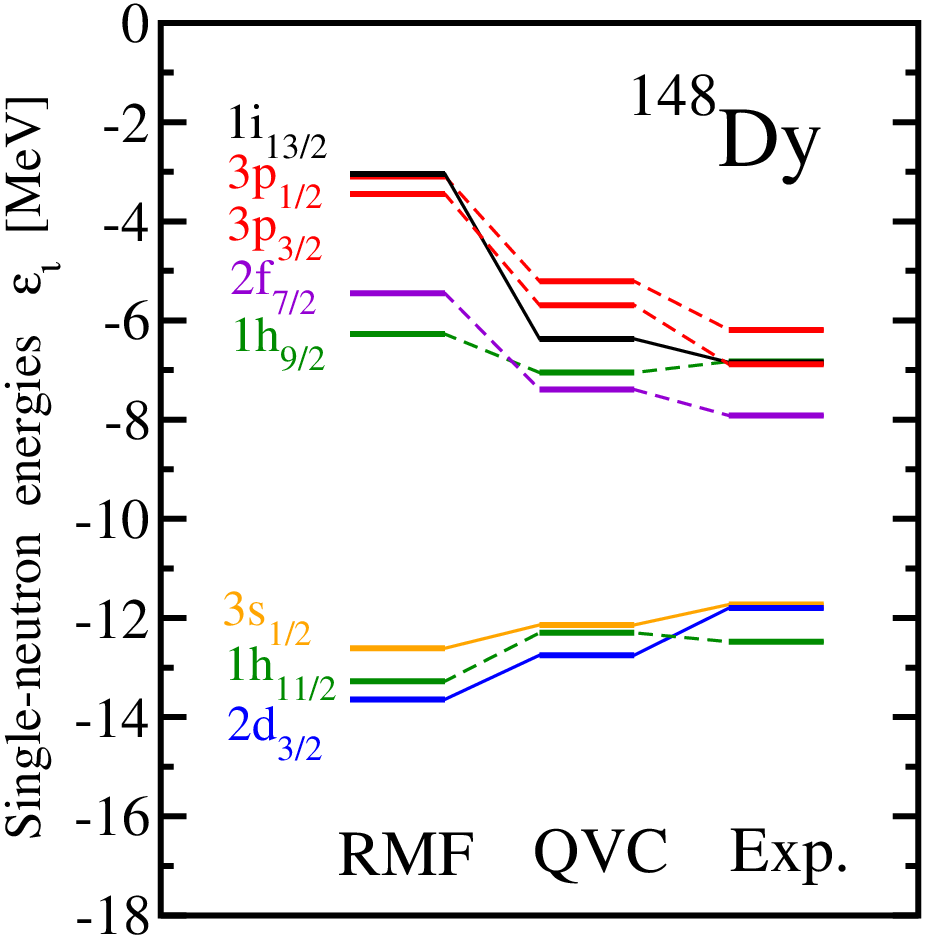}
\includegraphics[width=4.2cm]{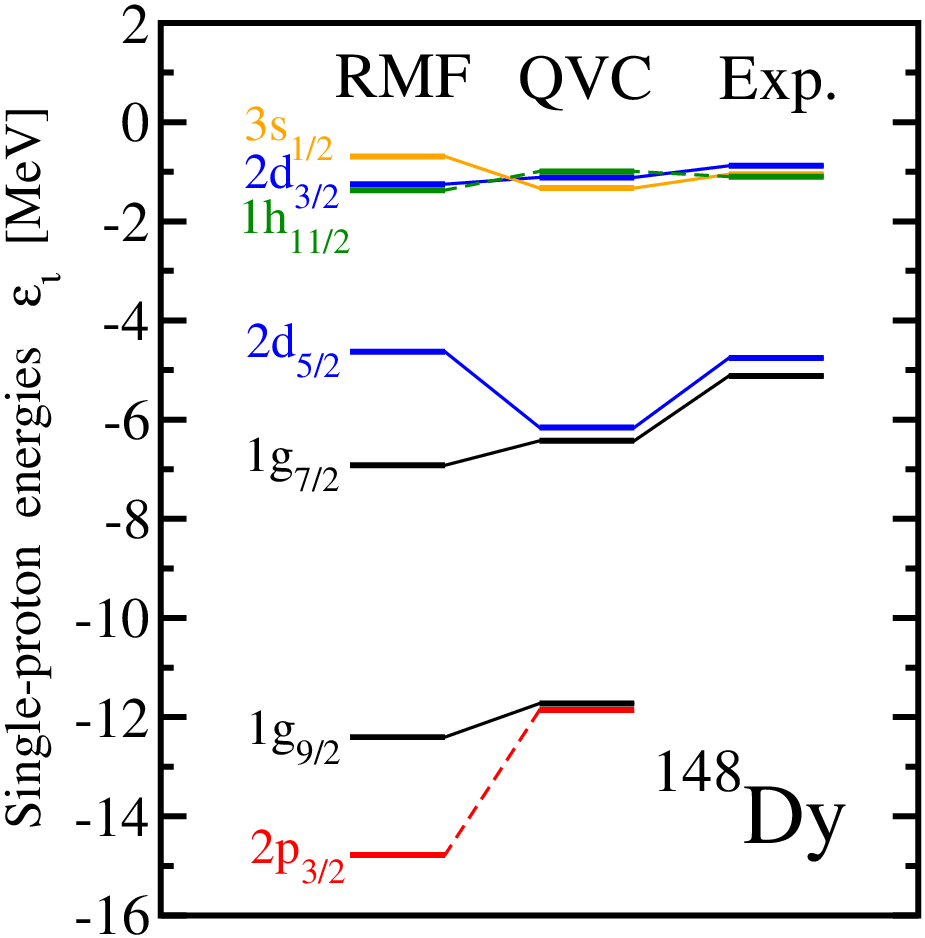}
\caption{(Color online) Spectra of $^{148}$Dy and its neighboring 
odd nuclei. Column ``RMF'' shows the single-quasiparticle spectra obtained 
in spherical RMF calculations of $^{148}$Dy. Column ``RQVC'' shows the 
spectra obtained in spherical calculations within the RQVC model
and $\Delta^{(5)}$ pairing. Column ``exp'' shows one-nucleon removal 
energies defined according to Eqs.\ (\ref{part_state}) and 
(\ref{hole_state}); they are based on the data of Refs.\ \cite{AME2012} 
(masses of ground states) and Ref.\ \cite{Eval-data-up} (the 
energies of excited states). Solid and dashed connecting lines between 
the states are used to indicate positive and negative parity states.}
\label{148Dy}
\end{figure}

\begin{figure}[ht]
\centering
\includegraphics[width=4.2cm]{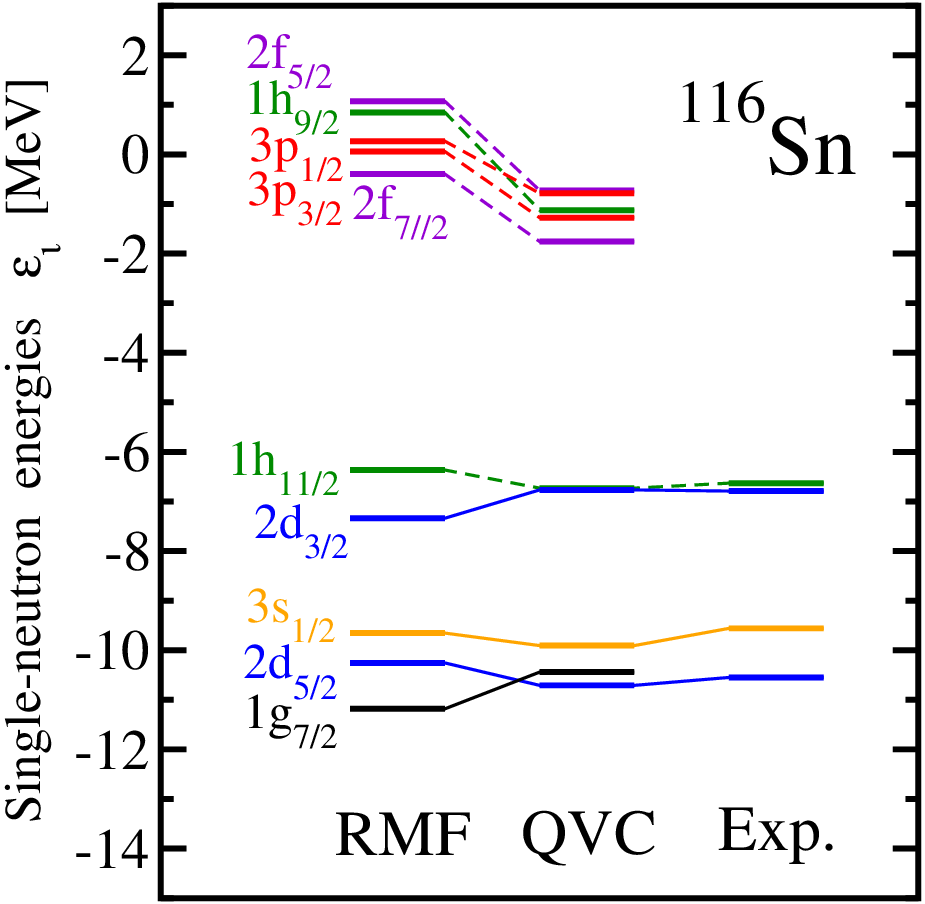}
\includegraphics[width=4.2cm]{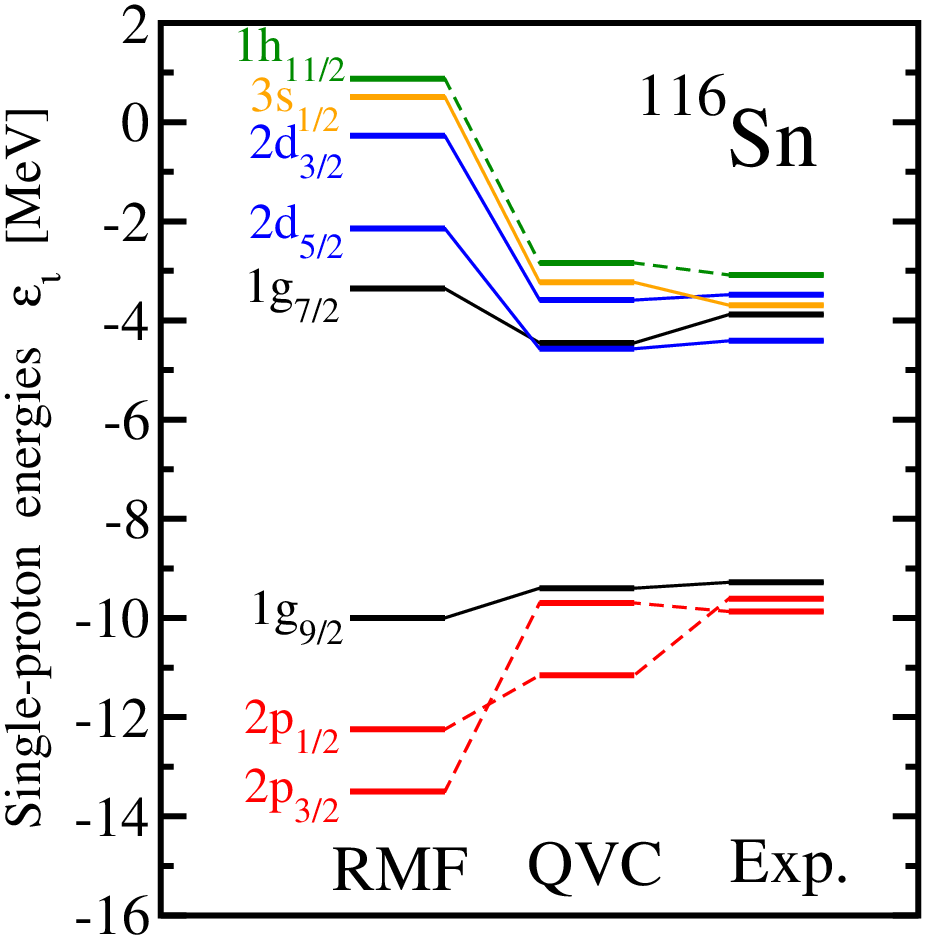}
\caption{(Color online) The same as in Fig.\ \ref{148Dy}, but for 
the spectra of $^{116}$Sn. }
\label{116Sn}
\end{figure}

The examples of the impact of quasiparticle-vibration coupling
on the spectra of odd-mass nuclei are shown in Figs.\ \ref{148Dy}
and \ref{116Sn}. In these figures, the column ``RMF'' shows the 
single-quasiparticle spectra obtained in spherical RMF calculations
of even-even nuclei. The dominant levels (i.e. the levels with the 
largest spectroscopic factors) as obtained in the RQVC 
calculations are shown in column ‘‘RQVC”. These results are 
compared with experimental energies of the dominant particle 
[$\varepsilon$(particle)] and hole [$\varepsilon$ε(hole)] 
states closest to the Fermi level determined from the difference 
of the binding energies of the even-even core [$B$(core)] and the 
corresponding adjacent odd nuclei [$B$(core + nucleon) and 
$B$(core - nucleon)] according to Refs.\ \cite{RBRMG.98,IEMSF.02} 
as
\begin{eqnarray}
\varepsilon {\rm (particle)} = B{\rm (core)}-B{\rm (core+nucleon)}
\label{part_state}
\end{eqnarray}
and
\begin{eqnarray}
\varepsilon {\rm (hole)} = B{\rm (core-nucleon)} - B{\rm (core)}.
\label{hole_state}
\end{eqnarray}
These quantities correspond to one-particle removal energies and 
they are shown in the column labeled as ``exp''.

 In Figs.\ \ref{148Dy} and \ref{116Sn} one can see the
two effects of the
quasiparticle-vibration coupling: (i) the general compression of the 
spectra leading to a substantially better agreement with experiment 
and (ii) in some cases the change of the level sequences. For some of 
the states, the QVC effect is quite substantial leading to a 2-3 MeV 
correction in the energy. These are, for example, neutron $1i_{13/2}$, 
$3p_{1/2}$ and $3p_{3/2}$ states in $^{148}$Dy and the proton $2d_{5/2}$, 
$2d_{3/2}$, $1h_{11/2}$ and $3s_{1/2}$ states in $^{116}$Sn. Similar large 
shifts in energy due to QVC can be found also for the states in proton 
subsystem of $^{148}$Dy and  neutron subsystem of $^{116}$Sn (see Fig.\ 
\ref{148Dy} and \ref{116Sn}).

 While the mean-field models deal with pure single-particle states,
in the RQVC model these states are fragmented due to the coupling to
vibrations, in some cases very strongly. However, around the 
Fermi surface there is often a dominant fragment with a considerable 
single-particle content. This is in agreement 
with numerous experimental observations. In this work we discuss only 
the dominant fragments of the single-quasiparticle states and show that 
the RQVC model reproduces their fractional occupancies.

 Figs.\ \ref{148Dy} and \ref{116Sn} also allow to understand the 
impact of the QVC on the energy splittings of the pairs of the 
dominant single-neutron $1i_{13/2}$ and  $1h_{9/2}$ states outside 
the $N=82$ core and of the dominant single-proton $1h_{11/2}$ and 
$1g_{7/2}$ states of the Sb isotopes outside $Z=50$ core. The evolution 
of these energy splittings along the isotonic/isotopic chains is 
discussed in details in Sec.\ \ref{splittings}. In each of
these pairs of the orbitals, one of the orbitals (neutron $1h_{9/2}$ 
in $^{148}$Dy and proton $1g_{7/2}$ orbital in $^{116}$Sn) is only
moderately affected by the QVC, while the other orbital (neutron $1h_{9/2}$ 
in $^{148}$Dy and proton $1g_{7/2}$ orbital in $^{116}$Sn) is substantially 
(by about 3 MeV) lowered by the quasiparticle-vibration coupling. As 
a result, the energy splittings between these pairs of the
orbitals are much smaller in the RQVC calculations as compared with 
mean field values.

\section{The energy splitting of the dominant single-particle 
         states sensitive to tensor interaction}
\label{splittings}

\begin{figure*}[ht]
\centering
\includegraphics[width=12.0cm]{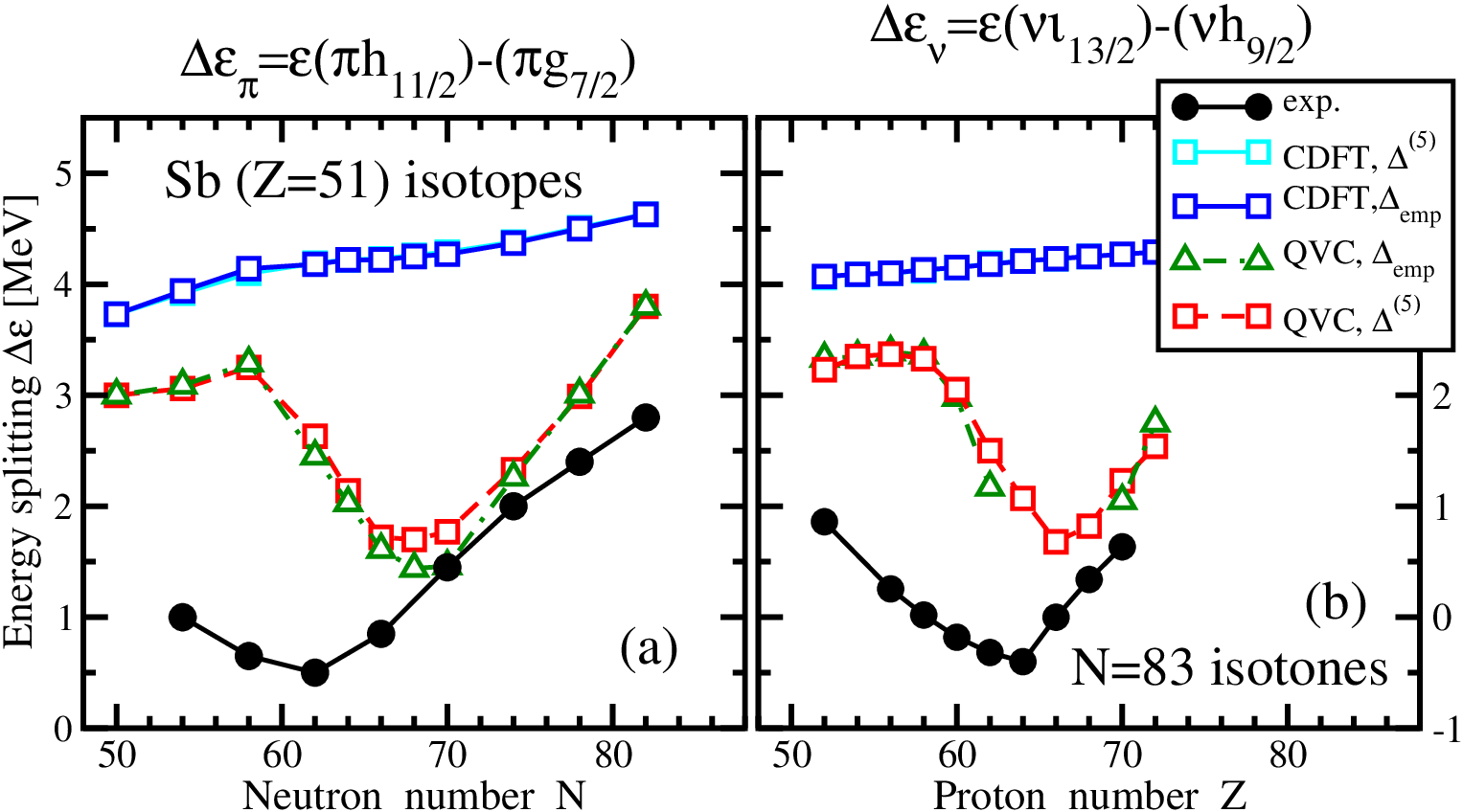}
\caption{(Color online) The energy splittings between the indicated
states obtained in experiment, covariant density functional theory
(CDFT) and RQVC calculations. The results of the calculations with
two pairing schemes are shown. Experimental data are taken from
Ref.\ \cite{Shif.04}.}
\label{splitting}
\end{figure*}

  In experiment, the energy splitting $\Delta \epsilon_{\pi} =
\epsilon(\pi h_{11/2}) - \epsilon(\pi g_{7/2})$ between the lowest
states of the Sb isotopes corresponding to the two nodeless
single-proton  $h_{11/2}$ and $g_{7/2}$ orbitals outside the closed
$Z=50$ core gradually decreases in the $N=54-62$ isotopes and then
rapidly increases for $N\geq 62$ (Fig.\ \ref{splitting}a).  The
change of the energy splitting $\Delta \epsilon_{\pi}$ on going from
$N=62$ to $N=82$ is substantial (around 2.5 MeV).  The CDFT
calculations (performed at the mean field level) give $\Delta
\epsilon_{\pi}\sim 4$ MeV which is smoothly increasing with neutron
number.

 One can see from Fig. \ref{splitting} that a large part of the 
corrections to the energy splittings and, in a half of the cases, the 
whole effect comes from the QVC. Indeed, the experimental $\Delta
\epsilon_{\pi}$ splitting and its trend with neutron number is
reasonably well reproduced in the RQVC calculations for the
$N\geq 68$ Sb nuclei. 
 Even in lighter nuclei the QVC
improves the agreement between theory and experiment as compared
with the CDFT results. 
Note that dependence on the pairing  gaps in the calculations 
for the $\Delta \epsilon_{\pi}$ and $\Delta
\epsilon_{\nu}$ splittings is relatively small (Fig.\ \ref{splitting}).
However, on going from $N=66$ to $N=50$ this 
improvement due to QVC is not sufficient
to fully reproduce experimental trend.
This is consistent with the behaviour of the energies of the first octupole 
states for small neutron numbers and also points out to the  increase of 
the importance of other mechanisms than those included into the RQVC
model. For example, isospin and pairing vibrations are usually not included 
in the RQVC phonon basis, however, their contribution can be important and 
should be studied in the future.

Unfortunately, the studies of the pion tensor coupling in
CDFT in Ref. \cite{LKSOR.09} and tensor interaction in Gogny DFT
in Ref.\ \cite{OSFGA.05} do not cover the Sb isotopes with $N=50-64$.
As a result, it is not clear whether the accounting of these
tensor couplings will improve the description of the $\Delta
\epsilon_{\pi}$  splittings for these nuclei. 
 The full $N=50-82$ range is covered only in Skyrme DFT
studies of Ref.\ \cite{CSFB.07} in which the tensor interaction 
is added to the Skyrme SLy5 functional. However, similar to our 
RQVC studies, the experimental $\Delta \epsilon_{\pi}$ splittings 
are described rather well only in $N\geq 70$ nuclei, and the 
discrepancy between theory and experiment is around 1 MeV for
lighter nuclei.

  A similar situation is seen in the energy splitting
$\Delta \epsilon_{\nu} = \epsilon(\nu i_{13/2}) - \epsilon(\nu h_{9/2})$
of the single-neutron $i_{13/2}$ and $h_{9/2}$ states outside the $N=82$
core (Fig.\ \ref{splitting}b). In the Skyrme DFT, this splitting is
reasonably well reproduced by the inclusion of the tensor interaction
\cite{CSFB.07} in high-$Z$ nuclei but the discrepancy between 
theory and experiment increases for $Z\leq 60$.
The CDFT mean-field calculations do not reproduce the
observed splittings,  but the accounting of the QVC improves the
description of the experimental $\Delta \epsilon_{\nu}$ values with
the biggest improvement seen in the middle of the shell (Fig.\
\ref{splitting}b). Similarly to the case of the Sb isotopes,
Figs.\ \ref{2+3-energy}b, d and \ref{splitting}b show that the best
agreement between experiment and theory for the $\Delta \epsilon_{\nu}$
is achieved for the nuclei in which the calculated excitation energies
of the  lowest $2^+$ and $3^-$ states are close to experimental ones.
Thus, the improvement in the description of these physical observables
should improve the description of the splittings.

  The comparison of our RQVC results with the DFT ones obtained
with non-relativistic Skyrme \cite{CSFB.07} and Gogny \cite{OMA.06}
functionals as well as relativistic \cite{LKSOR.09} functionals
indicates that both RQVC and tensor interaction act in the same
direction and reduce the  discrepancies between theory and experiment
for the $\Delta \epsilon_{\pi}$ and $\Delta \epsilon_{\nu}$ splittings.
If to consider both effects combined, the strong impact from RQVC on
the discussed splittings suggests that the effective tensor interaction
has to be quenched as compared with earlier estimates. 

\begin{figure*}[ht]
\centering
\includegraphics[width=13.5cm]{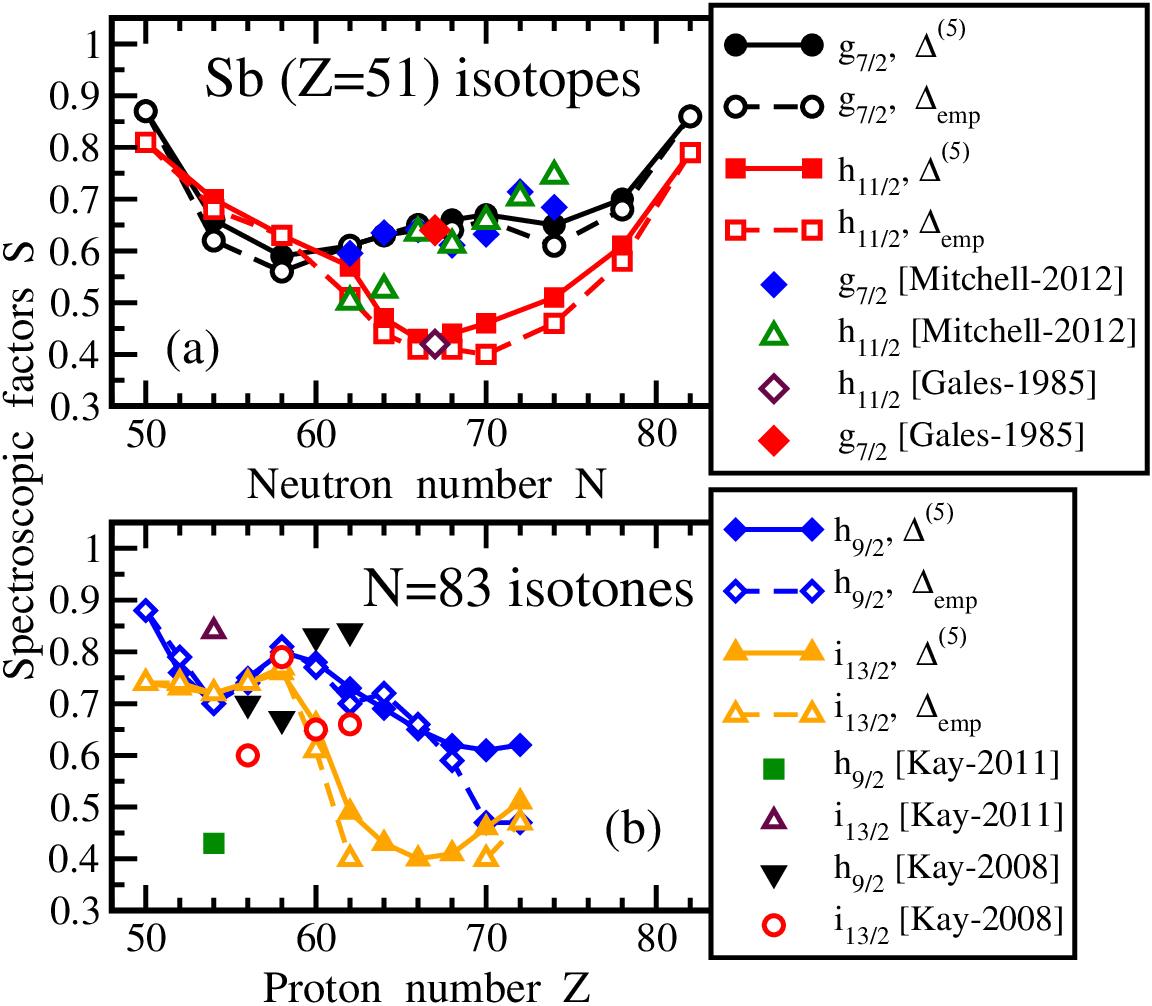}
\caption{(Color online) Spectroscopic factors $S$ of the states 
under study as functions of neutron and proton numbers.
Experimental data are taken from Refs.\ \cite{Mitch.12}
('Mitchell-2012'), \cite{117Sb-spec} ('Gales-2015'),
\cite{136Xe} ('Kay-2011') and \cite{Kay2008216}
('Kay-2008').}  
\label{s-factor}
\end{figure*}

\section{Spectroscopic factors of the states under consideration}
\label{s-factors}

Fig.\ \ref{s-factor} shows the spectroscopic factors $S$ of the dominant
states under investigation calculated within the QVC model and their
evolution with the particle number compared to available
data. One can see that near magic shell gaps these states are of predominant
single-particle nature ($S \sim 0.8$). However, the spectroscopic factors
$S$ of these states decrease on going away from the magic gaps indicating
the increased coupling of the single-particle motion with vibrations.
These trends correlate with the trends of the excitation energies of the
lowest $2^+$ and $3^-$ vibrational states. Indeed, the energies of these
vibrational states are  the highest for the doubly magic nuclei
and they are by a factor 2-3 smaller in open shell nuclei (see
Fig.\ \ref{2+3-energy}). As a result, the coupling with vibrations
is weaker in doubly magic nuclei and
stronger in open shell nuclei, which results in the reduced spectroscopic
factors $S$ of the considered dominant states in the nuclei away from the magic
ones. The calculated spectroscopic factors and the general trend of
their evolution with particle number (Fig.\ \ref{s-factor}) are
not very sensitive to the differences of the pairing gaps
$\Delta^{(5)}$ and $\Delta_{emp}$ .

It was deduced in Ref.\ \cite{Shif.04} that the states of interest
are almost single-particle in nature. Based on these results it
was concluded in the DFT framework that proper description of the
$\Delta \epsilon_{\pi}$ and  $\Delta \epsilon_{\nu}$ splittings
requires tensor interaction \cite{OSFGA.05,CSFB.07,LKSOR.09}.
Recently, an alternative analysis of Ref.\ \cite{Mitch.12} has
shown that the fragmentation of the $\pi g_{7/2}$ and $\pi h_{11/2}$
states is greater than in the study of Ref.\ \cite{Shif.04}.
These results support  the conclusions  of earlier analysis of
Refs.\ \cite{117Sb-spec,CHC.68,SP.08} which were in contradiction
with Ref.\ \cite{Shif.04}.  As compared to the experimental
data of Ref.\ \cite{Mitch.12}, our RQVC calculations reproduce well the
experimental spectroscopic factors of the $\pi g_{7/2}$ state in
all experimentally studied Sb nuclei and the ones of the $\pi h_{11/2}$
state in the $N=62,64$ Sb isotopes (Fig.\ \ref{s-factor}a)
and somewhat overestimate the fragmentation of the latter state in
the $N=66-74$ Sb isotopes. However, the $^{117}$Sb data on
the spectroscopic factors of the $\pi g_{7/2}$ and $\pi h_{11/2}$
states \cite{117Sb-spec} are well reproduced in the RQVC calculations 
(Fig.\ \ref{s-factor}a). In addition, in agreement with our RQVC
calculations the older experimental data of Ref.\ \cite{CHC.68}
(see analysis in Ref.\ \cite{SP.08}) provides low spectroscopic
factor $S\sim 0.5$ of the  $\pi h_{11/2}$ state in the mid-shell Sb
isotopes. Experimental spectroscopic factors depend considerably 
on the model analysis and on the reaction employed \cite{136Xe,LA.11};
this fact has also to be taken into account when calculated values
are compared with experimental ones. With this in mind one can also
conclude that apart of the $h_{9/2}$ state in $^{137}$Xe experimental 
spectroscopic factors of the $\nu i_{13/2}$ and $\nu h_{9/2}$  
states in the $N=83$ isotones are reasonably well reproduced in the RQVC 
calculations (Fig.\ \ref{s-factor}b). Observed strong fragmentation of the 
single-particle strength cannot be accounted for at the DFT level. This -
again weakens the conclusions of Refs.\ \cite{OSFGA.05,CSFB.07,LKSOR.09} 
on the need of strong tensor interaction.

 It the context of the current discussion it is interesting to mention
that the PVC calculations without pairing based on Skyrme functionals 
and restricted to odd-mass neighbors of doubly magic even-even nuclei 
point to either worsening (in $^{40}$Ca) or very limited (in $^{208}$Pb) 
improvement of the accuracy of the description of the spectra of such 
nuclei when tensor interaction is added \cite{CCSB.14}. Moreover, there 
are non-relativistic functionals without effective tensor interaction 
(such as SkM* and SkP) which provide better accuracy of the description 
of the $^{208}$Pb spectra 
(Ref.\  \cite{TDTC.14}) than the functional T44 with tensor interaction 
used in Ref.\ \cite{CCSB.14}. Note also that better accuracy than in T44
functional is achieved in the covariant PVC studies of Ref.\ \cite{LA.11} 
with the NL3* CEDF.

\section{Conclusions}
\label{concl}

  In conclusion, the impact of the quasiparticle-vibration coupling
on the energy splitting of specific pairs of the states in odd-mass
nuclei has been investigated in the relativistic framework. We
focus on the energy splittings between the proton $1h_{11/2}$ and
$1g_{7/2}$  states in the Sb ($Z=51$) isotopes and the neutron
$1i_{13/2}$ and $1h_{9/2}$ states in the $N=83$ isotones which,
according to the earlier studies within the DFT framework, can be
described only when effective tensor interaction is introduced.
Our analysis  unambiguously indicates that both QVC and tensor interaction
act in the same direction and reduce the discrepancies between
theory and experiment for the $\Delta \epsilon_{\pi}$ and 
$\Delta \epsilon_{\nu}$ splittings. This suggests that
the effective tensor force has to be quenched as compared with earlier 
estimates. These results also show that the definition of the strength 
of the tensor interaction by means of the 
fitting to the energies of the dominant single-quasiparticle states 
in odd-mass nuclei is flawed without accounting for the effects of 
QVC. For a quantitative analysis of the strength of effective 
tensor interaction in the density functional theories it is necessary 
to perform the calculations which incorporate both the quasiparticle-vibration 
coupling and tensor interaction  and, if possible, disentangle
explicitly their contributions.

\section{Acknowledgements}

 This material is based upon work supported by the U.S. Department of 
Energy, Office of Science, Office of Nuclear Physics under Award Number 
DE-SC0013037, by the U.S. National Science Foundation under the grants 
PHY-1204486 and PHY-1404343 and by the NSCL.

\bibliography{ref}

\end{document}